\documentclass[reprint,superscriptaddress,amsmath,amssymb,prl,floatfix]{revtex4-2}
\usepackage[dvipdfmx]{graphicx}
\usepackage{dcolumn}
\usepackage{bm}
\usepackage{xcolor}
\usepackage{here}

\usepackage{multirow}

\usepackage{amsthm}
\theoremstyle{definition}

\begin{document}
\title{Noise-induced order in high dimensions}

\author{Huayan Chen}
\affiliation{Department of Mathematics, Hokkaido University, N12 W7 Kita-ku, Sapporo, 0600812 Hokkaido, Japan}
\author{Yuzuru Sato}
\email{ysato@math.sci.hokudai.ac.jp}
\affiliation{Department of Mathematics, Hokkaido University, N12 W7 Kita-ku, Sapporo, 0600812 Hokkaido, Japan}
\affiliation{RIES, Hokkaido University, N12 W7 Kita-ku, Sapporo, 0600812 Hokkaido, Japan}
\affiliation{London Mathematical Laboratory, 14 Buckingham Street, London WC2N 6DF, United Kingdom}
\date{\today}

\begin{abstract}

Noise-induced phenomena in high-dimensional dynamical systems were investigated from a random dynamical system theoretical point of view. In a class of generalized Hénon maps, which are randomly perturbed delayed logistic maps, with monotonically increasing noise levels, we observed (i) an increase in the number of positive Lyapunov exponents from $4$ to $5$, and the emergence of characteristic periods at the same time, and (ii) a decrease in the number of positive Lyapunov exponents from $4$ to $3$, and an increase in Kolmogorov--Sinai entropy at the same time. Our results imply that simple concepts of noise-induced phenomena, such as noise-induced chaos and/or noise-induced order, may not describe those analogue in high dimensional dynamical systems, owing to coexistence of noise-induced chaos and noise-induced order.

\end{abstract}

\maketitle

            Noise-induced phenomena are caused by interactions between deterministic dynamics and external noise. When a transition occurs owing to small noise, the stationary distribution of the deterministic dynamical system is substantially altered, and the unobservable structure of the original dynamics becomes observable. In such cases, nonlinear phenomena, which qualitatively differ from deterministic dynamics, emerge in the noised dynamics. Typical examples of such noise-induced phenomena include stochastic resonance \cite{BENZI1982}, noise-induced synchronization \cite{Pikovskii1984, Teramae2004, carroll1995synchronizing}, noise-induced chaos (NIC) \cite{MayerKress1981, crutchfield1982fluctuations, tel2010quasipotential, RLM} and noise-induced order (NIO) \cite{Matsumoto1983, matsumoto1984noise}. 
            These low-dimensional noise-induced phenomena can be characterized by the Lyapunov spectrum, Kolmogorov--Sinai entropy, and power spectra. 
            In NIC, small noise from a periodic attractor induces chaotic behavior, resulting in the top Lyapunov exponent becoming positive, increasing the Kolmogorov--Sinai entropy, and causing the power spectrum to show broadband \cite{MayerKress1981}. In contrast, in NIO, small noise from a chaotic attractor induces ordered behavior, resulting in the top Lyapunov exponent becoming negative, reducing the Kolmogorov--Sinai entropy, and causing characteristic periods to emerge. Despite the importance of noise-induced transitions in scientific applications, explicit studies on noise-induced phenomena in high-dimensional dynamical systems is not well-developped except for a few works \cite{lai2003noise,shibata1999noiseless}. Tractable models for high dimensional dynamical systems have been developed for coupled dynamical systems \cite{Kuramoto1984, CML, Kaneko1989} and delayed dynamics \cite{Mackey1977,GHM}. Rössler studied high-dimensional chaotic atrractors that have more than one expanding direction \cite{Ross4d}, which can be characterized by the number of positive Lyapunov exponents, introduced here as $\kappa$. In low-dimensional NIC, the number of positive Lyapunov exponents, that is, the number of expanding directions in average, changes from $\kappa=0$ to $\kappa=1$ and in low-dimensional NIO, from $\kappa=1$ to $\kappa=0$, respectively. This scheme can be extended to high dimensional dynamical systems, allowing the changes in the number of positive Lyapunov exponents $\kappa$, in noised dynamics to be investigated. Indeed, we found multiple noise-induced transitions in random generalized Hénon maps, a class of high dimensional dynamical systems \cite{GHM}, changing $\kappa$ from $\kappa=4$ to $3$, $4$, and $5$, increasing the additive noise level monotonically.


        \begin{figure*}[htbp]
                \centering
                \includegraphics[width=0.75\textwidth]{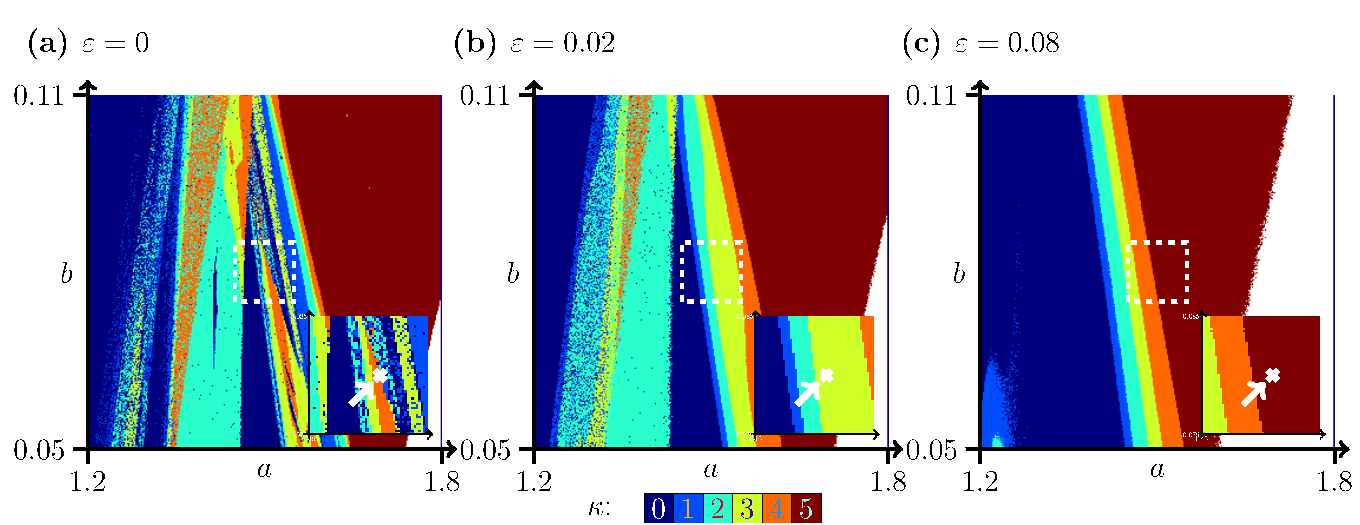}
                \caption{\label{fig:heatmap} 
                Phase diagram with the number of positive Lyapunov exponents $\kappa$ at $(a, b)$ shown in different colors for noise levels (a) $\varepsilon = 0$, (b) $\varepsilon = 0.02$, and (c) $\varepsilon = 0.08$. The white region indicates the divergence of dynamics. Numerical experiments have been done with $10^7$ iterations after $10^7$ step transients are ignored. Initial conditions are chosen randomly in $[-1.5, 1.7]^6$. The white cross represents $(a, b) = (1.51, 0.08)$, which is the model parameter we are investigating. The results indicate that the phenomenon of changing $\kappa$ is robustly observed.
                }
            \end{figure*}
            
        The generalized Hénon map is a delayed logistic map with $N-1$ time delay, which is given by;
    
        \begin{eqnarray} 
            \textbf{x}_{t+1} = \textbf{F}(\mathbf{x}_{t}), ~
            \mathbf{F}(\mathbf{x}_{t})= 
            \left [
            \begin{array}{ll}
            a-(x_{t}^{(N-1)})^2-bx_{t}^{(N)} \\
            x_{t}^{(1)} \\
            \vdots\\
            x_{t}^{(N-1)} \\
            \end{array}
            \right], ~ 
            \label{eq:GHM}
        \end{eqnarray}
        where ${\bf x}_t=(x^{(1)}_t,x^{(2)}_t,\ldots,x^{(N)}_t)$, $t = 0, 1, 2, \ldots$ is the iteration time, and $a,b$ are the parameters. The standard Hénon map recovers with $(a,b)=(1.4,-0.3)$ and $N = 2$ \cite{Henon1976}. Generalized Hénon maps may have up to $N-1$ positive Lyapunov exponents and show chaotic behavior \cite{RICHTER2002} (see supplemental materials on the bifurcations in generalized Hénon maps). We investigated the NIC with increasing $\kappa$, and NIO with decreasing $\kappa$ using the recently developed random dynamical systems theory. A random generalized Hénon map is given by 
        \begin{equation}
       \mathbf{x}_{t+1}=\mathbf{F}({\bf x}_t)+\varepsilon {\bf \Xi}_t, ~
       {\bf \Xi}_t=(\xi_t,0,\ldots,0)^T, 
       \end{equation}
       where $\xi_t \in [-1, 1]$ follows i.i.d. uniform distribution, and $\varepsilon$ is the parameter of the noise level. We adopted parameters $N = 6$, $a = 1.51$, and $b = 0.08$, and $\varepsilon \in [0, 0.1]$ as a control parameter in this paper.
        


            We present a phase diagram of the random generalized Hénon map with parameters $(a,b)\in [1.45, 1.55]\times [0.075, 0.085]$ by computing the Lyapunov spectra and observing the number of positive Lyapunov exponents $\kappa$ for the noise levels $\varepsilon=0, 0.02, 0.08$. In the case of $(a,b)=(1.51, 0.08)$, for the noiseless case with $\varepsilon=0$, we have $4$ positive Lyapunov exponents. For $\varepsilon=0.02$, we have $\kappa = 3$, and for $\varepsilon=0.08$, we have $\kappa= 5$. Because the Lyapunov exponents of random dynamical systems are typically continuous functions of the noise level, we initially observed decreasing $\kappa$ and then increasing $\kappa$ as the noise level increased. These multiple transitions have been reported in a class of one-dimensional maps, and its existence has been rigorously proved by computer assisted proofs \cite{galatolo2020existence, Chihara2022}. Positive measure regions were observed near $(a,b)=(1.51,0.08)$ in the phase diagram FIG. \ref{fig:heatmap} for all noise ranges, confirming that the phenomenon of changing $\kappa$ is robustly observed.

       
        \begin{figure}[htbp]
            \includegraphics[width=0.45\textwidth]{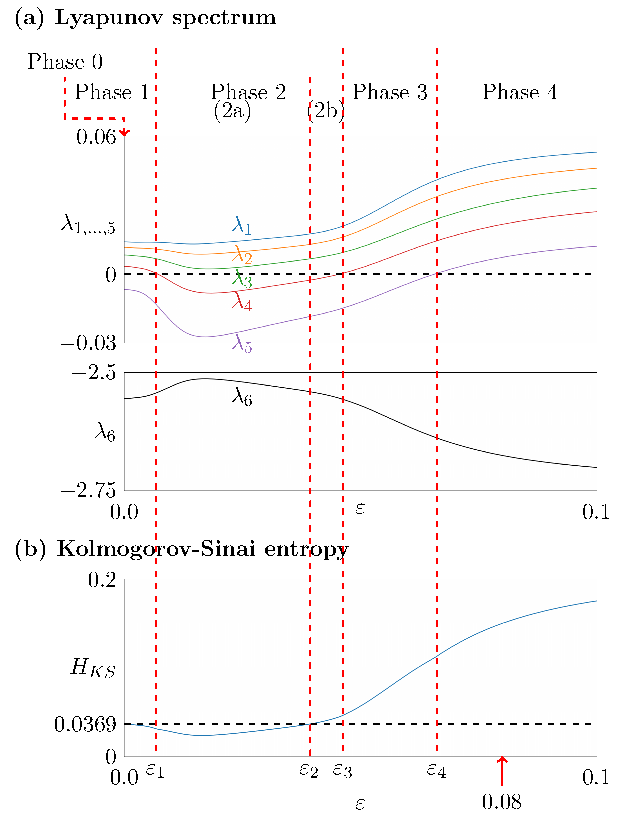}
            \caption{\label{fig:le_dky_hks}
            Illustrations showing the stability analysis in random generalized Hénon maps: (a) Lyapunov spectrum as a function of noise level $\varepsilon$; (b) Kolmogorov--Sinai entropy $H_{KS}$, which is estimated by the sum of positive Lyapunov exponents, as a function of noise level $\varepsilon$. Numerical experiments have been conducted with $10^9$ iterations after $10^7$ step transients are ignored, and the initial conditions are chosen randomly in $[-1.5, 1.7]^6$.}
        \end{figure}
        

        In Fig. \ref{fig:le_dky_hks}(a), the Lyapunov spectra $(\lambda_{1}, \lambda_2, \ldots,\lambda_{6})$ are shown. The range of the Lyapunov spectrum is shifted down and then shifted up. In Fig. \ref{fig:le_dky_hks}(b), the Kolmogorov--Sinai entropy $H_{KS}$ 
        is plotted as a function of the noise level $\varepsilon$. The Kolmogorov--Sinai entropy decreases first, and then increases. 
         We introduce $4$ different phases divided by the critical noise level, $\varepsilon =\varepsilon_1, \varepsilon_3, \varepsilon_4$. The critical noise levels are given by $\varepsilon_1 \approx 0.007$, $\varepsilon_2 \approx 0.0392$, $\varepsilon_3 \approx 0.0456$, $\varepsilon_4 \approx 0.0658$. 
         Phase $0$ is the deterministic limit with $\kappa=\kappa_0=4$ and $H_{KS} = h_0 \approx 0.0369$. In Phase $1 ~(0<\varepsilon <\varepsilon_1)$, $\kappa$ remains as $\kappa=\kappa_0$, however, $H_{KS} < h_0$, resulting in weak ordered behavior. Phase $2$ is divided into two sub-phases: $2a ~(\varepsilon_1 < \varepsilon <\varepsilon_2)$ and $2b ~ (\varepsilon_2 < \varepsilon < \varepsilon_3)$. In Phase $2a$, $\kappa=3 < \kappa_0$ and $H_{KS} < h_0$, showing typical NIO behavior. However, in Phase $2b$, $H_{KS} > h_0$, implying stronger chaoticity than Phase $2a$ while $\kappa$ actually decreases. In Phase $3 ~(\varepsilon_3< \varepsilon < \varepsilon_4)$, $\kappa = \kappa_0$, $H_{KS} > h_0$, which shows that the chaoticity is strengthened again. Finally, in Phase $4 ~(\varepsilon_4 < \varepsilon < \varepsilon_5)$, $\kappa=5 > \kappa_0$ and $H_{KS} > h_0$, showing typical NIC behavior.

\begin{figure}[htbp]
            \includegraphics[width=0.5\textwidth]{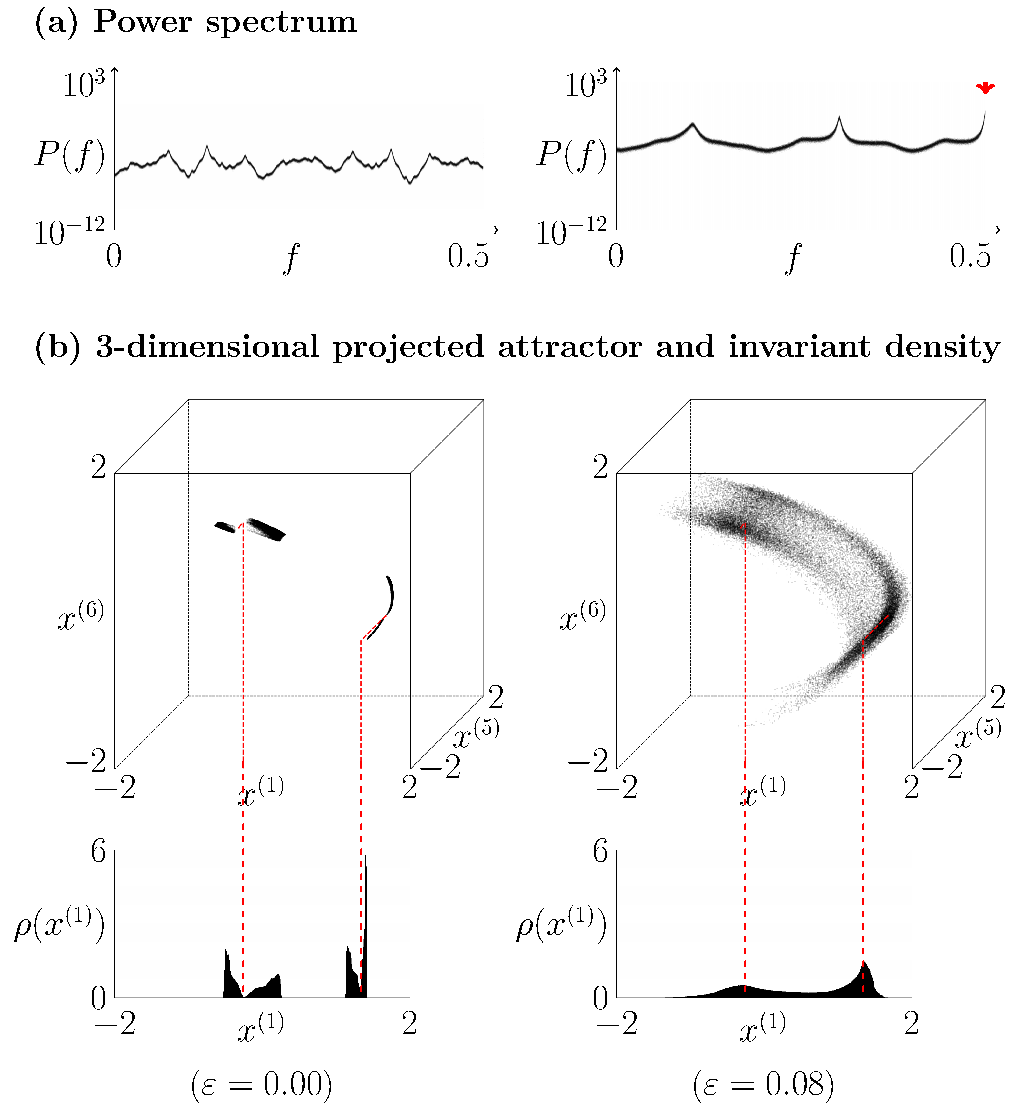}
                \caption{(a) Power spectrum $P(f)$ of the time series of $\{x^{(1)}_t\}$ as a function of $f$, with $\varepsilon = 0$ (left) and with $\varepsilon = 0.08$ (right). 
        (b) The attractor in the generalized Hénon map projected to a $3$-dimensional sub-space $(x^{(1)},x^{(5)},x^{(6)})$ (top) and the corresponding invariant densities $\rho(x^{(1)})$ projected to a $1$-dimensional sub-space $x^{(1)}$ (bottom), with $\varepsilon = 0$ (left) and $\varepsilon = 0.08$ (right).
        The UPO $P_2$ is depicted by red points and red dashed lines. By adding noise, band merging happens (see supplemental materials on the band merging) to form a global stretching-and-folding structure. The random attractor includes $P_2$, and a clear characteristic period of $2$ emerges. Numerical experiments have been done similarly as in Fig. \ref{fig:le_dky_hks}.}
                \label{fig:geometry}
            \end{figure}

        Next, to discuss the attractor geometry and periodicity, we particularly focused on a period-2 and a period-4 unstable periodic orbit (UPO) $P_2: \{{\bf q}_{1}, {\bf q}_{2}\}$ and a period-4 unstable periodic orbit $P_4: \{{\bf p}_{1}, {\bf p}_{2}, {\bf p}_3, {\bf p}_4\}$ 
        in the deterministic generalized Hénon map, where
        
            \begin{eqnarray}
            && \mathbf{q}_{1} \approx (-0.257,  1.337, -0.257,  1.337, -0.257,  1.337), \nonumber \\ 
            && \mathbf{q}_{2} \approx ( 1.337, -0.257,  1.337, -0.257,  1.337, -0.257), \nonumber
            \end{eqnarray}  
            and 
            \begin{eqnarray}
            && \mathbf{p}_{1} \approx (-0.456,  1.398,  0.128,  1.191, -0.456,  1.398), \nonumber \\ 
            && \mathbf{p}_{2} \approx ( 1.398,  0.128,  1.191, -0.456,  1.398,  0.128), \nonumber \\
            && \mathbf{p}_{3} \approx ( 0.128,  1.191, -0.456,  1.398,  0.128,  1.191), \nonumber \\ 
            && \mathbf{p}_{4} \approx ( 1.191, -0.456,  1.398,  0.128,  1.191, -0.456), \nonumber
            \end{eqnarray} 
            which are computed by Newton's method. We expected that, with the addition of noise, these UPOs would attract orbits and enhance the stability and periodicity of the noised dynamics. Fig. \ref{fig:geometry}(a) illustrates the power spectrum and the characteristic peak as indicated by a red arrow. Fig. \ref{fig:geometry}(b) shows attractors projected to a $3$-dimensional sub-space $(x^{(1)},x^{(5)},x^{(6)})$ (top), and the corresponding invariant densities $\rho(x^{(1)})$ projected to a $1$-dimensional sub-space $x^{(1)}$ (bottom). The UPO $P_2$ is shown through red points, and its projection is indicated by red dashed lines.

           In Fig. \ref{fig:geometry} (a) (left), no prominent peaks are observed in the power spectrum of the deterministic time series; however, in Fig. \ref{fig:geometry} (a) (right), we observed a clear characteristic peak at the frequency $f \approx 1/2$, implying that the characteristic period is $2$. Fig. \ref{fig:geometry}(b)(left) shows a deterministic attractor with $\kappa_0=4$ positive Lyapunov exponents. This attractor is a $4$-band chaotic attractor, and the UPO $P_2$ is located outside of it. In contrast, the random attractor presented in Fig. \ref{fig:geometry}(b)(right) is a single-band chaotic attractor, including the UPO $P_2$. Its invariant density concentrates near $P_2$, implying that the noised orbits concentrate to the UPO $P_2$. Simultaneously, the attractor represents a global stretching-and-folding structure, which contributes to the increase in $\kappa=5>\kappa_0$.

         Table 1 summarizes our observationd in the numerical experiments for the numbers of positive Lyapunov exponents $\kappa$, Kolmogorov--Sinai entropy $H_{KS}$, concentration of invariant density, peaks in power spectrum, the numbers of major clusters (effectively the number of observable bands), and stretching-and-folding structure. According to these statistics and geometrical structure, we expect that there are mixed states with coexistence of NIC and NIO.  For instance, with $\varepsilon = 0.04$ in Phase 2, $\kappa$ decreases from $4$ to $3$, which indicates stabilizationm, however, the Kolmogorov--Sinai entropy increases, indicating increase in uncertainty of the dynamics. Similarly, in Phase 4, $\kappa$ increases from $4$ to $5$, which is also shown by the emergence of a global stretching-folding structure, indicating stronger chaoticity, while the peak at $f=1/2$ in the power spectrum indicates the newly emerging peirodicity of the noised dynamics. Similar phenomena with the UPO $P_4$ are observed in the case of weak noise (see supplemental materials on the periodicity with the UPO $P_4$.). In these cases, it is thought that the UPOs such as $P_2$ and $P_4$ attract the noised orbits, the peaks in the invariant density emerges, and the weak periodicities are observed.

        \begin{table}[htbp]
            \centering
                \begin{tabular}{l|c|c|c|c|c|c}
               
                \hline
                Phase & 
                        0 & 
                        1 & 
                        2a &  
                        2b &  
                        3 &  
                        4 \\
                \hline
                 \hline
                 $\varepsilon  $& 
                        $ 0 $ & 
                        $(0, \varepsilon_1)$ &  
                        $(\varepsilon_1, \varepsilon_2)$ &   
                        $(\varepsilon_2,  \varepsilon_3)$ &   
                        $(\varepsilon_3,  \varepsilon_4)$ &   
                        $(\varepsilon_4,  0.1)$ \\
                \hline
                 $\kappa  $& 
                        4& 
                        4&  
                        3&
                        {\color{blue}3}&   
                        4&   
                        {\color{red}5}\\
                \hline
                $H_{KS}$ &
                        $ = h_0 $ &
                        $ < h_0$ &
                        $ < h_0$ &
                        {\color{blue}$ > h_0$} &
                        $ > h_0$ &
                        $ > h_0$ \\
                \hline
                Density & & & & & & \\
                concentration&
                        - &
                        $P_4$&
                        $P_4$&
                        $P_4, P_2$&
                        $P_2$&
                        {$P_2$}\\
                \hline
                Characteristic& & & & & & \\
                frequency &
                        - &
                        $ {1/4 }$&
                        $ {1/4}$&
                        $ {1/4, 1/2 }$&
                        $ {1/2 }$&
                        {\color{red}$ {1/2}$}\\

                \hline
                Number of& & & &\multicolumn{2}{c|}{} & \\
                major clusters&
                        4 &
                        4 $\to$ 2 &
                        2 &
                        \multicolumn{2}{c|}{2 $\to$ 1} &
                        1 \\
                \hline
                Stretch\&fold&
                        -&
                        local&
                        local&
                        \multicolumn{2}{c|}{local $\to$ global} &
                        global\\
                \hline
              
                \end{tabular}
            \caption{\label{tab:summary} 
              Summary of the numerical analysis for random generalized Hénon maps with $N = 6$ and $(a, b) = (1.51, 0.08)$. The critical noise levels are given by $\varepsilon_1 \approx 0.007$, $\varepsilon_2 \approx 0.0392$, $\varepsilon_3 \approx 0.0456$, $\varepsilon_4 \approx 0.0658$. 
In Phase $0$, $h_0\approx 0.0369$ is the Kolmogorov--Sinai entropy without noise. NIC with enhanced regularities is indicated in red. NIO with enhanced uncertainty is indicated in blue.}
        \end{table}

        This paper reports on noise-induced phenomena in 6-dimensional random generalized Hénon maps, which are randomly perturbed delayed logistic maps. At monotonically increasing noise levels, we observed (i) an increase in the number of positive Lyapunov exponents from 4 to 5 and the emergence of characteristic periods at the same time, and (ii) a decrease in the number of positive Lyapunov exponents from 4 to 3 and an increase in the Kolmogorov--Sinai entropy at the same time. Our results imply that simple concepts of noise-induced phenomena, such as NIC and NIO, may not describe those analogue in high dimensions, owing to the mixed states with coexistence of NIC and NIO. In presented examples of the random generalized Hénon maps, NIC with enhanced regularity and NIO with enhanced uncertainty are observed. Further random dynamical system analysis for these heterogeneous noise-induced transitions will be studied elsewhere. Noise-induced phenomena in continuous flows, which are described by stochastic differential equations (see supplemental materials on NIC in extended Rössler systems), are promissing future works to carry out.

\vspace{5mm}
\begin{acknowledgements}
    Y.S. was supported by JSPS Grant-in-Aid for Scientific Research (B), JP No. 21H01002. 
\end{acknowledgements}

\bibliography{thesis_ref}

\end{document}